\documentclass[a4paper,pra,preprint,floatfix,superscriptaddress]{revtex4}
\usepackage[english]{babel}
\usepackage{amsmath,amsfonts,amssymb,float}
\usepackage{graphicx}
\usepackage{dcolumn,bm}

\def\clf{Central Laser Facility, STFC Rutherford Appleton Laboratory, Didcot, OX11 0QX, United Kingdom}
\def\strathclyde{SUPA, Department of Physics, University of Strathclyde, Glasgow, G4 0NG, United Kingdom}

\def\Z{\mathbb{Z}}
\def\Q{\mathbb{Q}}

\begin{document}

\title{Laser harmonic generation with tuneable orbital angular momentum using a structured plasma target}

\author{R. M. G. M. Trines}
\author{H. Schmitz}
\affiliation\clf
\author{M. King}
\author{P. McKenna}
\affiliation\strathclyde
\author{R. Bingham}
\affiliation\clf
\affiliation\strathclyde
\date\today

\begin{abstract}

In previous studies of spin-to-orbital angular momentum (AM)
conversion in laser high harmonic generation (HHG) using a plasma
target, one unit of spin AM is always converted into precisely one
unit of OAM \cite{zepf,li2020}. Here we show, through analytic
theory and numerical simulations, that we can exchange one unit of SAM
for a tuneable amount of OAM per harmonic step, via the use of a
structured plasma target. In the process, we introduce a novel
framework to study laser harmonic generation via recasting it as a
beat wave process. This framework enables us to easily calculate and
visualise harmonic progressions, unify the ``photon counting'' and
``symmetry-based'' approaches to HHG and provide new explanations
for existing HHG results. Our framework also includes a specific way
to analyse simultaneously the frequency, spin and OAM content of the
harmonic radiation which provides enhanced insight into this
process. The prospects of using our new framework to design HHG
configurations with tuneable high-order transverse modes, also
covering the design of structured plasma targets, will be discussed.

\end{abstract}

\maketitle

\section{Introduction}

The generation of high harmonic radiation (HHG) with higher order phase or polarisation topology, especially Laguerre-Gaussian (LG) modes with orbital angular momentum (OAM) has attracted strong interest recently. This involves both HHG in gases using pump beams that already have OAM \cite{gariepy,garcia13, rego16,rego19, torusknot,rego22} and HHG using CP beams without OAM but exploiting spin-to-orbital momentum conservation in laser-solid interactions \cite{zhang15,zepf,li2020,zhang21} or laser-aperture interactions \cite{GonzalezNatPhys2016, GonzalezNatComm2016, DuffSciReports2020, yi21, Jirka2021, Bacon2022}. The harmonic generation is usually assumed to follow these principles: (i) no energy, linear or angular momentum can be left behind in the medium or target, (ii) HHG in an isotropic medium (gas) conserves spin and orbital angular momenta separately within the EM waves, so HHG by pure CP waves is not possible as it would violate spin conservation, (iii) HHG in laser-solid interaction conserves total angular momentum within the EM waves, so a single CP wave without OAM can produce CP harmonics with OAM via spin-to-orbital angular momentum conversion  \cite{zepf,li2020}. In this paper, we aim to go beyond these principles: when a laser pulse interacts with a target with a special surface structure, angular momentum can be left behind in the target and the OAM content of the harmonics can differ significantly from the prediction of the simple SAM-OAM conversion. In the process, we will also introduce a systematic approach to simultaneous HHG in multiple variables (e.g. frequency $\omega$ and OAM level $\ell$) and novel diagnostic methods that provide more insight into the multi-dimensional harmonic progression.

\section{Conventions}

In this paper, we study high harmonics generated by a cubic nonlinear
term $|\Psi^2| \Psi$ in the wave equation for $\Psi$. This covers the most
common scenarios for HHG. Throughout this paper, we will use the
following conventions.
\begin{enumerate}
  \item We write all laser fields as superpositions of circularly polarised laser modes. For example, a laser pulse with linear (elliptic) polarisation is the superposition of \emph{two} CP modes with equal (unequal) amplitudes. 
  \item We use the function $\Psi = A_x + i A_y$ to describe the transverse EM fields, where $(A_x,A_y)$ represents the transverse vector potential. For a CP vector potential $(\cos(\omega t - \ell \varphi),\sigma \sin(\omega t - \ell \varphi))$, where $\sigma = \pm 1$ denotes the spin direction, we find that $\Psi = \exp i\sigma (\omega t - \ell \varphi)$. For two such modes $\vec{A}$ and $\vec{B}$, we find that $\vec{A}\cdot\vec{B} = \Re(\Psi_B^* \Psi_A) \propto \cos[(\sigma_B \omega_B - \sigma_A \omega_A)t - (\sigma_B \ell_B - \sigma_A \ell_A)\varphi]$. This convention will also allow us to make use of the complex Mathieu equation.
  \item In light of the above, we will use the products $\sigma \omega$, $\sigma \vec{k}$, $\sigma \ell$, etc. instead of $\sigma$ and $\omega$ separately. While $\sigma\omega$ can be any real number, we will always use $\omega > 0$, which uniquely defines $\sigma$. This convention renders both the calculation and the visualisation of harmonic progressions easier and more intuitive.
  \item We will treat left and right circularly polarised modes at the same harmonic frequency $\omega$ as separate harmonics with opposite signs of $\sigma\omega$ rather than different polarisations of the same harmonic. We do this so we can treat the spin $\sigma$ on a footing equal to the frequency $\omega$, wave vector $\vec{k}$ or OAM level $\ell$. For example, sending a left CP laser beam through a half-wave plate to create a right CP beam will be treated as generating harmonic $-1$, since $\sigma\omega \to -\sigma\omega$ in that case.
  \item In terms of the harmonic progression, we treat harmonic cascades in quantities like $\omega$, $k_\perp$ or $\ell$ on an equal footing as much as possible. So a cascade in $\ell$ at a single frequency $\omega_0$ \cite{vieira2} is treated as a harmonic progression just like a cascade in $\omega$ at the single OAM level $\ell=0$. This way, multi-dimensional cascades in, for example, $(\sigma\omega,\sigma\ell)$ space are also possible.
\end{enumerate}

The justification for these conventions is as follows. (i) A CP wave has constant amplitude $A^2$ and does not beat with itself (unlike an LP wave), while two CP waves with parallel $\vec{k}$ beat in precisely one way. (ii) A CP wave is a spin eigenstate, while LP and EP waves are not. Thus, CP waves can be used to verify the spin conservation law, while LP waves cannot be used for that. (iii) A CP wave has constant $A^2$, so obeys all possible translation/rotation symmetries. Adding a second CP wave will break the symmetry of $A^2$ in one specific way, leading to high harmonic generation in one specific dimension in $(\omega\vec{k})$ space. This relates the dimensionality of symmetry breaking to the dimensionality of the HHG progression. Thus, CP waves are more fundamental ``building blocks'' to study HHG than LP waves.

Using the above conventions renders it easier to calculate and visualise the harmonic progression. The harmonic progression becomes a ``beat wave'' process \cite{beat1,beat2}: the beating fixes the harmonic step to $\sigma_B \omega_B - \sigma_A \omega_A$ or $\sigma_B \ell_B - \sigma_A \ell_A$ etc., while $\sigma\omega$ (or $\sigma\ell$ etc.) can assume any positive or negative value. It is easy to extend this procedure to more variables ($k_\perp$, $\ell$, etc.) or to more CP modes. The harmonic spectrum generated by two CP modes will be a sequence of equidistant points on a line; for three CP modes that are not on the same line in 2-D Fourier space, the generated harmonic spectrum will be a 2-D grid; for four CP modes that are not on the same plane in 3-D Fourier space, the generated harmonic spectrum will be a 3-D lattice.

\section{Harmonic progression}
\label{sec:3}

In this section, we will study the harmonic progression for several common nonlinear systems. In nearly all cases, the lowest-order nonlinear term is either cubic, e.g. $|\vec{A}|^2 \vec{A}$, or quadratic, e.g. $2(\vec{A}\cdot\vec{B})\vec{A}$ or similar, where $\vec{B}$ is a vector constant in time. In general, the cubic case is seen in HHG in isotropic media where no energy or momentum is left behind in the medium, so energy and momentum are conserved in the EM waves alone. The quadratic case is often seen in HHG in anisotropic media where linear or angular momentum (or even energy, as in stimulated Raman scattering) can be left behind in the medium. In the quadratic case, the nonlinear terms can often be rewritten as $|\vec{A}+\vec{B}|^2 \vec{A}$ or even $|\vec{A}+\vec{B}|^2 (\vec{A}+\vec{B})$, effectively reducing it to a cubic case providing a joint description of the laser and the medium, see also Section \ref{sec:9}. For this reason, we first study the case of a wave equation with cubic nonlinearity (we will discuss more complex nonlinear terms later):
\begin{equation}
\label{eq:model}
(\partial_t^2 - c^2 \nabla^2) \Psi = -(\Omega^2  + \Gamma^2 \Psi^* \Psi) \Psi.
\end{equation}

We use a pump beam $\Psi_0 = \Psi_A + \Psi_B + \Psi_C + \ldots$, where $\Psi_A = a_A \exp i\sigma_A (\omega_A t - k_A z - \ell_A \varphi - k_{\perp,A} x_\perp - \delta_A)$, etc. For a single mode $\Psi_0 = \Psi_A$, we find that $\Psi_0^* \Psi_0$ is constant, and no harmonics will be generated. For two modes, $\Psi_0 = \Psi_A + \Psi_B$, we can use the theory of the Mathieu equation and the Jacobi-Anger expansion to find the harmonic frequency progression:
\begin{equation}
\label{eq:spinenergy}
\sigma_n \omega_n = \sigma_A \omega_A + n(\sigma_B \omega_B - \sigma_A \omega_A),\quad n \in \Z.
\end{equation}
Since $\omega_n > 0$ always, $\sigma_n$ is well-defined. We note that all the frequencies in the $\sigma\omega$ spectrum are equidistant, while this need not be true at all for the $\omega$ spectrum.

Now that $\sigma_n$ has been determined, we introduce the other harmonic progressions ($n \in \Z$):
\begin{align}
\label{eq:spinoam}
\sigma_n \ell_n &= \sigma_A \ell_A + n(\sigma_B \ell_B - \sigma_A \ell_A),\\
\sigma_n k_{\perp,n} &= \sigma_A k_{\perp,A} + n(\sigma_B k_{\perp,B} - \sigma_A k_{\perp,A}),\\
\sigma_n \delta_n &= \sigma_A \delta_A + n(\sigma_B \delta_B - \sigma_A \delta_A - \arg(\Gamma^2)),\\
\sigma_n^2 &= \sigma_A^2 + n(\sigma_B^2 - \sigma_A^2),
\end{align}
where the last equation is trivially true since $\sigma^2 = 1$ always.

We define $\#_A$, $\#_B$, etc. as the number of photons absorbed from mode A, B, etc. for the generation of mode $n$. Assuming energy conservation \cite{bloembergen80,schafer93,corkum93,nonper2}, i.e. $\omega_n = \#_A \omega_A + \#_B \omega_B$, we find:
\begin{equation}
\label{eq:photonnumber}
(\#_A,\#_B)_n = \sigma_n [(\sigma_A,0) + n(-\sigma_A,\sigma_B)].
\end{equation}
We note that every quantity that scales in the same way as $\omega_n$, i.e. $\sigma_n$, $\ell_n$, and $k_{\perp,n}$, will be automatically conserved when $\omega_n$ is conserved, due to the form of the expressions for $\#_A$ and $\#_B$. Thanks to our conventions, we can calculate $\#_A$ and $\#_B$ explicitly, and get conservation laws for spin, orbital angular momentum and transverse momentum for free.

The above harmonic progression is characterised by one index $n$ and is essentially one-dimensional. However, with three independent CP modes, a two-dimensional progression in two variables, e.g. $\omega$ and $\ell$, can be realized. The harmonic progression then becomes:
\begin{align}
\sigma_{nm} \omega_{nm} &= \sigma_A \omega_A + n(\sigma_B \omega_B - \sigma_A \omega_A) + m(\sigma_C \omega_C - \sigma_A \omega_A),\\
\sigma_{nm} \ell_{nm} &= \sigma_A \ell_A + n(\sigma_B \ell_B - \sigma_A \ell_A) + m(\sigma_C \ell_C - \sigma_A \ell_A).
\end{align}
The most visible difference between the 1-D and 2-D progressions is that in a 1-D progression there will always be exactly one OAM level for a given harmonic frequency, while in a 2-D progression there can often be many OAM levels for a given harmonic frequency. This will be expanded on below.

The perturbative expression for the harmonic amplitude is then approximately given by:
\begin{equation}
\label{eq:pert}
    a_n \propto \prod_i \frac{a_i^{|\#_i|}}{|\#_i|!}.
\end{equation}
From this we see that the order of the amplitude $a_n$ is given by $\sum_i |\#_i|$, while the order of the harmonic $n$ is given by $|\sum_i \#_i|$. From Eq. (\ref{eq:photonnumber}), we find that $\sigma_A \#_A + \sigma_B \#_B = \sigma_n$, so $|\sigma_A \#_A + \sigma_B \#_B|=1$. For $\sigma_A \sigma_B \#_A \#_B > 0$, we find that $|\#_A|+|\#_B|$ remains constant. Likewise, for three or more modes, we can prove that $\sigma_A \#_A + \sigma_B \#_B = \sigma_n - \sigma_C \#_C$, etc. For only two modes A and B, one will always have $\sigma_A \sigma_B \#_A \#_B < 0$, and $|\#_A|+|\#_B|$ is not conserved under a step (A,B). This means that the perturbative harmonic amplitude, Eq. (\ref{eq:pert}), will always decay exponentially with photon and/or harmonic number. In turn, this implies that harmonics involving large numbers of photons are ``mathematically possible'', but only the lowest few of those have a large enough amplitude to be ``physically relevant''. This has prompted numerous authors to look for non-perturbative scenarios of HHG where the amplitude decay with harmonic number is polynomial or power-law \cite{nonper1, nonper2, long95, nonper3}, usually up to a certain cutoff. 

For three or more modes A, B, C, \ldots, there is a subset of the harmonic spectrum where $\sigma_A \sigma_B \#_A \#_B > 0$, so $|\#_A|+|\#_B|$ is preserved under a step (A,B). This means that a step (A,B) will not change the order $|\#_A|+|\#_B|+\ldots$ of the harmonic amplitude in the subset of the harmonic spectrum where $\sigma_A \sigma_B \#_A \#_B > 0$. In other words, the harmonics generated by the beating term $\Psi_A^* \Psi_B$ will have a certain range where their amplitude, Eq. (\ref{eq:pert}), shows polynomial scaling rather than exponential, with the cutoff defined by $\sigma_A \sigma_B \#_A \#_B = 0$ (see also the ``binomial'' scaling of Rego \emph{et al.} \cite{rego16}). However, non-perturbative models may well predict a wider range of harmonics generated by $\Psi_A^* \Psi_B$ with polynomial or power-law scaling, i.e. ``physically relevant'' amplitudes even for $\sigma_A \sigma_B \#_A \#_B < 0$, see e.g. Rego \emph{et al.} \cite{rego16}.


\section{HHG in laser-solid interaction: the non-linear q-plate}
\label{sec:9}




\subsection{Wave equation}

High harmonic generation (HHG) using circularly polarised beams to generate
circularly polarised harmonics has attracted a lot of interest recently.
This is most easily accomplished via the use of two pump beams with
counter-rotating circular polarisations in a gas target
\cite{long95,fleischer,kfir2}, since this
conserves spin angular momentum in a natural way. In laser-solid
interactions, HHG using a single laser beam with circular polarisation
has been achieved if either the laser incidence was non-perpendicular
\cite{lichters} or the target was non-planar
\cite{koba12,zepf, li2020, yi21, zhang21}
(See also Huang \emph{et al.} \cite{huang20}.)
In particular, if the laser beam interacts with a circular indentation
or aperture \cite{koba12,zepf}, the EM wave cannot exchange angular
momentum with the
target, and the total angular momentum (AM) of the EM waves is then conserved
during the HHG process: if $n$ photons with circular polarisation at the
fundamental frequency $\omega$, each carrying 1 unit of spin AM, combine
to produce one circularly polarised photon at $n\omega$, this harmonic
photon will carry 1 unit of spin AM and $n-1$ units of orbital AM to
arrive at a total AM of $n$ units.

In the next few sections, we will investigate HHG with circularly polarised
beams where the target does not have circular symmetry, so the total AM is
\emph{not} conserved in the EM waves alone during the HHG process. For this,
we take inspiration from the so-called $q$-plate \cite{qplate1,qplate2}:
an optic that flips
the polarisation of circularly polarised light and gives it orbital
AM in the process, absorbing 2 units of AM and emitting $2q$ units.
The total AM of the beam then changes by $2(q-1)$, which is nonzero
for $q\not= 1$. In HHG, this would mean that the harmonic $n\omega$
carries 1 unit of spin AM and $nq - 1$ units of orbital AM, where
$q$ is determined by the specific shape of the target.

For HHG in laser-solid interaction, we use the model by Lichters \emph{et al.}
\cite{lichters}, which describes the interaction of a laser beam with
a solid surface. The action of the laser beam makes the surface (and thus
the reflection point of the laser light) oscillate with either frequency $2\omega$
(perpendicular incidence or oblique incidence with s-polarisation)
or frequency $\omega$ (oblique incidence with p-polarisation). For oblique
incidence at an angle $\alpha$ with respect to the target normal, when the
projection of $\vec{k}$ on the target surface is given by $k\sin\alpha
\vec{e}_x$, the model equations are as follows ($\vec{a}_{DC} =
\tan\alpha \vec{e}_x$):
\begin{align}
(\partial_t^2 -c^2 \nabla^2) (\vec{a} -\vec{a}_{DC})
&= -\omega_p^2 \frac{n}{\gamma\cos\alpha}(\vec{a} - \vec{a}_{DC}),\\
n/\gamma &\approx 1 + \delta n - (\vec{a} - \vec{a}_{DC})^2/2,\\
\partial_t^2 \delta n &= -\frac{c^2}{2\gamma^2} \nabla^2 (\vec{a} - \vec{a}_{DC})^2,\\
\gamma^2 &= 1 + (\vec{a} - \vec{a}_{DC})^2.
\end{align}

We define $\Psi = (a_x - \tan\alpha) + ia_y$ and only retain leading-order
terms for small wave amplitudes. Then we can rewrite the above system as:
\begin{equation}
\cos\alpha(\partial_t^2 -c^2 \nabla^2) \Psi + \omega_p^2 \Psi =
-\frac12 \omega_p^2 (1+c^2 k^2/\omega^2) (\Psi^* \Psi) \Psi,
\end{equation}
which has the required shape, Eq. (\ref{eq:model}). However, this requires that
$\Psi$ does not just cover the EM vector potential $\vec{a}$, but also a
synthetic ``direct current (DC) mode'' $\vec{a}_{DC} = \tan\alpha \vec{e}_x$,
which is a consequence of the laser-target geometry. The presence of this
DC mode explains why
a single laser beam with circular polarisation can generate harmonics when
hitting a target at an oblique angle \cite{lichters}: unlike in HHG in gas,
the circularly polarised mode is not on its own but can beat against the
DC mode.

The DC mode for oblique laser incidence onto a planar target has
$\Psi_{DC} = \tan\alpha =\mathrm{Const}$, which means that $\sigma\omega
= \sigma k = \sigma\ell = 0$. It should also be regarded as having linear
polarisation in $x$. With this, judicious application of (\ref{eq:spinenergy})
will yield all the selection rules described in Ref. \cite{lichters}.
While their beams did not have OAM, a pump
beam with oblique incidence, OAM level $\ell$ and circular polarisation
would have generated harmonics with OAM levels $\sigma_n \ell_n =
n(\sigma\ell)$.

The same strategy with DC modes can be used to describe e.g. second-harmonic
generation in nonlinear materials, see e.g. Ref. \cite{maki95}. The cubic
term in the equation for $\vec{a}$ can be rewritten as either $(\vec{a}-\vec{a}_1)^2 (\vec{a}-\vec{a}_1)$
or $(\vec{a}-\vec{a}_1)^2 (\vec{a}-\vec{a}_2)$ or $[(\vec{a}-\vec{a}_1)
\cdot (\vec{a}-\vec{a}_2)] (\vec{a}-\vec{a}_3)$. If one then equips the
DC modes with the right higher-order structure (e.g. OAM), the second
harmonic light can then be generated with intrinsic OAM levels or other
higher order structure.

To proceed: a more generic way to describe a laser pulse at oblique incidence
on a target is to use a height function $h(x,y)$ to describe the target surface,
and define $\vec{a}_{DC} = \nabla_\perp h$ or $\Psi_{DC} = \partial h/\partial
x + i\partial h/\partial y$. For
oblique incidence on a plane target this yields $h(x,y) = x\tan\alpha$
and $\Psi_{DC} = \tan\alpha$. For more intricate surface shapes, there are
almost unlimited possibilities for $\Psi_{DC}$.

Wang \emph{et al.} \cite{zepf} studied the interaction of a circularly
polarised laser beam, $\sigma=+1$, with a conically shaped dent in the
target surface, half angle $\alpha$. Li \emph{et al.} \cite{li2020} studied
the interaction of a tightly focused laser pulse having a concave phase
front with a planar target, which is sufficiently similar to Ref. \cite{zepf}
for our purposes. The height function for a conical dent is $h(x,y) =
\tan(\pi/2-\alpha) \sqrt{x^2 + y^2}$ so $\Psi_{DC} = \tan(\pi/2-\alpha)
\exp(i\varphi)$, where $\varphi$ is the azimuthal angle in polar
coordinates. Then the DC mode has $\sigma\omega = 0$ and $\sigma \ell = -1$.
If this dent is hit by a circularly polarised laser beam with spin $\sigma_0$,
frequency $\omega_0$ and OAM level $\ell_0$, then we find from (\ref{eq:spinenergy})
that $\sigma_n \omega_n = n(\sigma_0\omega_0)$ and
$\sigma_n \ell_n = n(\sigma_0\ell_0 + 1)-1$ for all $n\in\Z$. These results 
encompass the selection rules provided by Refs. \cite{zepf,li2020}, and also
include the ``negative frequencies'' for $n<0$, which have been predicted
by Li \emph{et al.} \cite{li2020} but not yet demonstrated.

Combining all of the above, we wish to obtain a laser-plasma interaction
that generates high harmonics satisfying ($n\in\Z$, $q\in\Q$):
\begin{align}
\label{eq:qfreq}
\sigma_n\omega_n &= n\sigma\omega,\\
\label{eq:qoam}
\sigma_n\ell_n &= n(\sigma\ell + q) -q.
\end{align}
This generalises all of the above. Setting $q=0$ returns the results
by Lichters \emph{et al.} \cite{lichters}. Setting $q=1$ returns the
results from
Refs. \cite{zepf,li2020}. Setting $n=-1$ and thus $\sigma_n=-\sigma$
yields an ordinary $q$-plate \cite{qplate1,qplate2}. We note
that Eqns. (\ref{eq:qfreq}) and (\ref{eq:qoam}) correspond to Eqns.
(\ref{eq:spinenergy}) and (\ref{eq:spinoam}) in Section \ref{sec:3}
with $\omega_A = 0$ and $\sigma_A \ell_A = -q$. 

\subsection{Shape of the target surface}

To obtain a nonlinear $q$-plate, we need to create
a DC mode with $\Psi_{DC} \propto \exp(iq\varphi)$, i.e. a shaped
target surface with a height function such that $\nabla_\perp h \propto
(\cos(q\varphi),\sin(q\varphi))$. To determine a suitable $h(x,y)$, we
first identify its contour lines, i.e. the curves $(r(\varphi),\varphi)$
in polar coordinates for which the polar angle $\alpha$ of the tangent
to the curve satisfies $\alpha(\varphi) = q\varphi + \alpha_0$.
This leads to the following equation:
\[
r'(\varphi)/r(\varphi) = \cos[\alpha(\varphi) 
-\varphi]/\sin[\alpha(\varphi) -\varphi].
\]
For $q \not= 1$, we can set $\alpha_0 = \pi/2$ without loss of
generality, and we find $(C>0)$:
\begin{equation}
\label{eq:contour1}
r(\varphi) = C[\cos(q-1)\varphi]^{1/(q-1)}.
\end{equation}
For $q=1$ and $\alpha_0 \not= 0$, we find:
\begin{equation}
\label{eq:contour2}
r(\varphi) = C\exp(a\varphi),\ a = \cot(\alpha_0).
\end{equation}
For $q=1$ and $\alpha_0=0$, we find: $\varphi = C$, $r$ undetermined.

The next step is to determine the height function $h(x,y)$ or $h(r,\varphi)$
that we need in the laser-target interaction where we generate
our harmonics. In Ref. \cite{zepf}, a cone-shaped dent is used with
height function $h(r,\varphi) = r\tan(\pi/2-\alpha)$, to
obtain laser-plasma HHG with $q=1$. The contour lines for this
shape satisfy: $z=C$, $r(\varphi) = C$, that is, Eq. (\ref{eq:contour2})
with $q=1$ and $\alpha=\pi/2$. This can easily be generalised. For $q\not= 1$,
we set $h(r,\varphi) = C^{1-q}$ and invert Eq. (\ref{eq:contour1}) to obtain:
\[
h(r,\varphi) = r^{1-q} \cos[(q-1)\varphi].
\]
For $q>1$, this leads to a pole at $r=0$, so we may need to exclude
the area $r<\epsilon$ for some small $\epsilon>0$, or use e.g.
$h(r,\varphi) =\arctan\{ r^{1-q} \cos[(q-1)\varphi] \}$ to saturate
$h$ near $r=0$. This is not necessarily
a problem, since laser beams with nonzero OAM usually have vanishing
intensity for $r \downarrow 0$, so the precise shape of the target surface
near $r=0$ is not too important. As an alternative,
one can set $h = C^{q-1}$ when $q>1$ and invert Eq. (\ref{eq:contour1})
to obtain $h(r,\varphi) = r^{q-1}/ \cos[(q-1)\varphi]$, and saturate $h$
near the angles where $\cos[(q-1)\varphi] = 0$.

For $q=1$ and $\alpha=\pi/2$, we recover the conical shape of Ref.
\cite{zepf}. For $q=1$ and $\alpha=0$, we set $h(r,\varphi) = \lambda_0
\varphi/(2\pi)$ for $r \geq \epsilon > 0$ to obtain the ``spiral
plate'' of Ref. \cite{li2020}. For $q=0$, we obtain $h(r,\varphi) =
r\cos(\varphi) = x$, which corresponds to a laser beam with oblique
incidence onto a plane, as described by Lichters \emph{et al.}
\cite{lichters}. We see that we can design
target surfaces to obtain the harmonic generation patterns dictated
by Eqns. (\ref{eq:qfreq}) and (\ref{eq:qoam}) for any $q$.




The concept of the non-linear $q$-plate can be extended to that of
the non-linear J-plate \cite{jplate1,jplate2}. This works as follows. A basic
J-plate turns a circularly polarised beam with $\sigma=+1$ into one
with $\sigma=-1$, adding $r$ to the OAM level. Similarly, it turns a
circularly polarised beam with $\sigma=-1$ into $\sigma=+1$, adding
$s$ to the OAM level. To ensure that $r$ and $s$ are fully independent,
we need two degrees of freedom in the design of the non-linear
$q$-plate, which we choose to be $q$ and $\ell_s$. We revisit
Eq. (\ref{eq:qoam}), with nonzero $\ell_s$:
\[
\sigma_n\ell_n = n[\sigma(\ell+\ell_s) + q] -q.
\]
We set $n=-1$ and use $\sigma_{-1} = -\sigma$ to obtain:
\[
\ell_{-1} = \ell+\ell_s + 2\sigma q
\]
For $\sigma=+1$, we find $\ell_{-1} - \ell = r = \ell_s + 2q$, while
for $\sigma=-1$, we find $\ell_{-1} - \ell = s = \ell_s - 2q$. Thus,
a non-linear J-plate with parameters $r,s$ can be produced from
an enhanced non-linear $q$-plate by setting $q = (r-s)/4$ and
$\ell_s = (r+s)/2$.

Note that $\nabla_\perp h$ of a non-linear $q$-plate corresponds to
the ``frozen'' field of a Laguerre-Gaussian mode with circular polarisation
and OAM level $\sigma\ell = -q$, see e.g. Courtial \emph{et al.}
\cite{courtial98}. We can exploit this to design target surfaces
that correspond to e.g. the frozen field of Hermite-Gaussian modes.
Irradiate this with a Hermite-Gaussian mode with linear polarisation
parallel to $\nabla_\perp h$, to obtain odd harmonics with the HG
mode of the pump, even harmonics with the HG mode of the target. There
are many more possibilities, e.g. also exploring the radial mode of
the Laguerre-Gaussian harmonics, but that is beyond the scope of this
paper.

\subsection{Aperture targets}

In addition to the reflective targets with a structured surface discussed above, we
have also considered aperture targets where the laser beam passes through an
aperture in a solid target. The inner surface of the aperture may be purely circular
or may have a corrugated structure. Harmonics are generated during the interaction
of the laser light with the inner aperture surface, and both the pump laser and the
harmonic radiation are diagnosed behind the target. The concept of the \textit{relativistic
plasma aperture} was introduced, and explored experimentally and numerically, in
Gonzalez-Izquierdo \textit{et al} \cite{GonzalezNatPhys2016}, in which it was shown that the collective
plasma electron dynamics can be controlled by variation of the polarisation of the drive laser 
pulse. It was also demonstrated that modulations in the fast electron distribution are mapped into the spatial-intensity profile of the beam of accelerated ions, via modulation of the electric field generated, and subsequently the ion beam profile is influenced by the laser polarisation \cite{GonzalezNatComm2016}. In subsequent experiments and simulations to investigate predicted changes to the polarisation of the laser light transmitted through the relativistic plasma aperture it was discovered that intense laser light in the fundamental and second harmonic is generated in high order modes that can evolve on intra-pulse time-scales \cite{DuffSciReports2020}. The mode structure and polarisation state vary with the interaction parameters and result from a superposition of coherent radiation, generated by a directly accelerated bipolar electron distribution, and the light transmitted due to the onset of relativistic self-induced transparency. The possibility exists to develop this approach to achieve dynamic control of structured light fields at ultrahigh intensities. It has also been shown that the diffracted light that travels though the aperture contains high-harmonics of the fundamental laser frequency \cite{yi21}. In addition, 3D particle-in-cell simulations by Jirka \textit{et al} \cite{Jirka2021} indicate that the laser intensity may be increased roughly by an order of magnitude in the near field behind a relativistic plasma aperture. Whereas, all of these studies involve a self-generated relativistic plasma aperture in an initially opaque ultrathin foil target, Bacon \textit{et al} \cite{Bacon2022} shows that the same physics, resulting in second-harmonic laser light with higher-order spatial modes, occurs when starting with a preformed aperture target. It is further shown that it is possible to change between a linearly polarized TEM01 mode and a circularly polarized Laguerre-Gaussian LG01 mode by changing the polarisation of the drive laser pulse, enabling selectable spatial mode structure.

In the case of the reflective targets above, the HHG process was governed by the term $\vec{E}\cdot\nabla h$, where the topology of the vector field $\nabla h$ determines the harmonic progression in $(\sigma\omega,\sigma\ell)$ space. In the case of an aperture target, the role of $\nabla h$ is taken over by the inner normal vector $\vec{n}$ of the aperture, and the ``beating term'' $\vec{A}\cdot\nabla h$ is replaced by $\vec{A}\cdot\vec{n}$ \cite{yi21}. With this identification, we will be able to design targets to produce the same harmonic progressions, whether the target is ``aperture'' or ``reflective''.

\section{Design of numerical simulations and data analysis}

\subsection{Simulation setup}


\emph{Reflective target simulations}
We perform simulations of a short intense laser pulse impacting a solid density target and the subsequent high harmonic generation using the EPOCH Particle-in-Cell code \cite{Arber:2015}. The simulation domain is given by $-8\mu\mathrm{m} \le x, y, z \le 8\mu\mathrm{m}$ which is resolved by $512 \times 512 \times 1024$ grid cells. This means the resolution is double in the $z$--direction compared to the transverse $x$ and $y$--directions. The simulation is set up with a perfect vacuum in the left half of the simulation box and a constant plasma density in the right half of the simulation box. The vacuum-plasma interface is given by the height function $z_{\mathrm{int}} = h(x, y)$ such that $n(z < z_{\mathrm{int}}) = 0$ and $n(z \ge z_{\mathrm{int}}) = 4 n_c$. Here $n_c$ is the critical density calculated for a laser wavelength of $1\mu\mathrm{m}$. Only electrons are simulated; the ions are treated as an immobile background. Where the electron density is non-zero, we initialise the simulation with 5 particles per cell. A short high-intensity laser pulse is injected from $z = -8\mu\mathrm{m}$ boundary propagating in the positive $z$--direction. The pulse has a Gaussian profile with a half width of $w_0 = 4\mu\mathrm{m}$ and a Gaussian temporal envelope with a duration of 100fs. The wavelength is $\lambda=1\mu\mathrm{m}$ and the peak intensity is $I_{\mathrm{max}} = 6.636 \times 10^{18} \mathrm{W cm}^{-2}$ which corresponds to an $a_0 \approx 2.2$. The boundaries are periodic in the $x$ and $y$--directions, and a perfectly matched layer absorbs the reflected laser pulse in the negative $z$-direction.

At each time-step during the simulation, we write out the electromagnetic fields in a fixed plane $z_0=-7\mu\mathrm{m}$ to obtain the time dependent fields $\mathbf{E}(x, y, z_0, t)$ and $\mathbf{B}(x, y, z_0, t)$. These fields serve as the basis for our analysis. We integrate the fields over the radial direction from $r=0$ to $1\mu\mathrm{m}$ for a constant azimuthal angle $\phi$  to obtain $\mathbf{E}(\phi, t)$ and $\mathbf{B}(\phi, t)$. In order to separate out the forward and backward propagating waves, we make use of the fact that, for a plane wave in vacuum with wave vector $\mathbf{k}$ the vectors $\mathbf{E}$, $\mathbf{B}$ and $\mathbf{k}$ form a right handed system, i.e. $E_y = c B_z $ and $E_z = - c B_y$ for waves propagating in the positive $z$--direction, and $E_y = -c B_z $ and $E_z = c B_y$ for waves propagating in the negative $z$--direction. Using these properties, we define the complex amplitudes
\begin{align*}
a^{\mathrm{fw}}(\phi, t) &= E_y + c B_z  + i(E_z - c B_y)\\
a^{\mathrm{bw}}(\phi, t) &= E_y - c B_z  + i(E_z + c B_y)
\end{align*}
for the forward and backward propagating waves respectively. We analyse the modes in the reflected laser pulse by calculating the Fourier transform of $a^{\mathrm{bw}}$ to obtain $\hat{a}^{\mathrm{bw}}(N_{\phi}, \omega)$. The complex amplitudes allow us to distinguish left-hand circularly polarised modes, $\omega < 0$ from right-hand polarised modes $\omega > 0$.

\emph{Aperture simulations}
To investigate the harmonic generation from aperture targets, 3D particle-in-cell (PIC) simulations were conducted using the EPOCH code [ref]. These consisted of an [x,y,z] spatial domain of $[\pm10 \lambda_L,\pm10 \lambda_L,-5\ldots 15\lambda_L]$ with 720 $\times$ 720 $\times$ 1000 simulation cells and all boundaries defined as free-space. The targets were composed of a slab of Al$^{13+}$ ions of 
thickness $\lambda_L$ with an empty circular aperture with a periodically varying radius, $R$. This is defined as $R=R_0-R_1$cos$(n\phi+\pi)$ where $R_0$=3.75$\lambda_L$,
$R_1$=0.75$\lambda_L$, $\phi=$atan2$(x,y)$ and n is the number of periods. This was neutralised with an electron population with a peak density equal to 40n$_{crit}$ and an initial temperature of 10 keV. The front surface of the 
target was set at z=0. Each species was initialised with 80 particles per cell. 
The laser pulse was circularly polarised ($\sigma_0=1, l_0=0$) in the [x,y] direction and focused with a Gaussian spatial profile at z=0 resulting in a full-width half-maximum (FWHM) of 3.5$\lambda_L$. The temporal profile was also defined 
as Gaussian with a FWHM of 12$\tau_L$. The peak intensity was 10$^{21}$ Wcm$^{-2}$

\subsection{Diagnosing the harmonics}
\label{sec:10}

In this section, we discuss how the harmonic modes in our simulations can be diagnosed properly. This is especially important in case of a circularly polarised pump and a harmonic with opposite spin to the pump, since such harmonics tend to be drowned out by their counterpart with the same spin as the pump. See e.g. Li \emph{et al.} \cite{li2020}, where harmonics with opposite
spin are predicted but not observed, as their amplitude is assumed to be ``too small'' to be detected easily.

A transverse real-valued electric field $\vec{E}_\perp$ can be decomposed into two orthogonal linear polarisations: $\vec{E}_\perp = (\vec{E}_\perp \cdot \vec{e}_x)\vec{e}_x + (\vec{E}_\perp \cdot \vec{e}_y)\vec{e}_y$, where $\vec{E}_\perp \cdot \vec{e}_x$ and $\vec{E}_\perp \cdot \vec{e}_y$ are independent real-valued quantities. Alternatively, $\vec{E}_\perp$ can also be decomposed into two orthogonal circular polarisations: $\vec{E}_\perp = E_+\vec{e}_+ + E_-\vec{e}_-$, with $\vec{e}_\pm = (\vec{e}_x \pm i\vec{e}_y)/\sqrt{2}$ and $E_\pm = \vec{E}_\perp \cdot \vec{e}_\pm$ complex-valued. However, since $\vec{E}_\perp^* = \vec{E}_\perp$ while $\vec{e}_-^* = \vec{e}_+$, we find that $E_- = E_+^*$. In the frequency domain, we find that $\mathcal{F}[E_-](\omega) = (\mathcal{F}[E_+](-\omega))^*$, i.e. the RCP frequency spectrum for $\omega > 0$ is the complex conjugate of the LCP frequency spectrum for $\omega < 0$ and vice versa. This motivates us to (i) only use the $E_+$ coefficient, or actually $\Psi = E_+\sqrt{2}$, and (ii) use the ``signed frequency'' $-\infty < \sigma\omega < \infty$, so $\mathcal{F}[E_+](\sigma\omega)$ captures all the spectral information of $\vec{E}_\perp$, while each value of $\sigma\omega$ represents a mode with pure circular polarisation.

Furthermore, we recall that a ``fundamental'' CP mode is represented by $\Psi = \exp i\sigma\theta$, where $\theta = \omega t - k_z z - k_\perp x_\perp - \ell\varphi - \ldots$. This means that the use of $\sigma\omega$ in the frequency domain also implies the use of $\sigma k$, $\sigma\ell$, etc. in the momentum domain, rather than simply $k$ or $\ell$. We will use this later when determining two-dimensional Fourier spectra, e.g. the two-dimensional Fourier transform from $(t,\varphi)$ space to $(\sigma\omega,\sigma\ell)$ space to diagnose the joint frequency-OAM spectrum.


For the case of a non-linear $q$-plate, we apply a two-dimensional
Fourier analysis, from $(z,\varphi)$-space to
$(\sigma k_z,\sigma \ell)$-space. From the real-valued $E$-field of
our simulations, we calculate the complex function $\Psi = E_x + i E_y$.
Next, we take the two-dimensional fast Fourier transform of
$\Psi(z,\varphi)$ and re-center to obtain $\tilde\Psi(\sigma k_z,
\sigma \ell)$ for $-\infty < \sigma k_z < \infty$, and use $\sigma\omega
\sim c\sigma k_z$. This will separate L- and R-polarised modes for
each frequency, and reveal linear relationships like
$\sigma_n \ell_n + 1 = (\sigma\ell + 1)
\sigma_n \omega_n / (\sigma\omega)$, or Eq. (\ref{eq:spinenergy}).

\section{Simulation results}

\subsection{Reflective target simulations}

\begin{figure}[ht]
    \begin{center}
    \includegraphics[width=0.3\columnwidth]{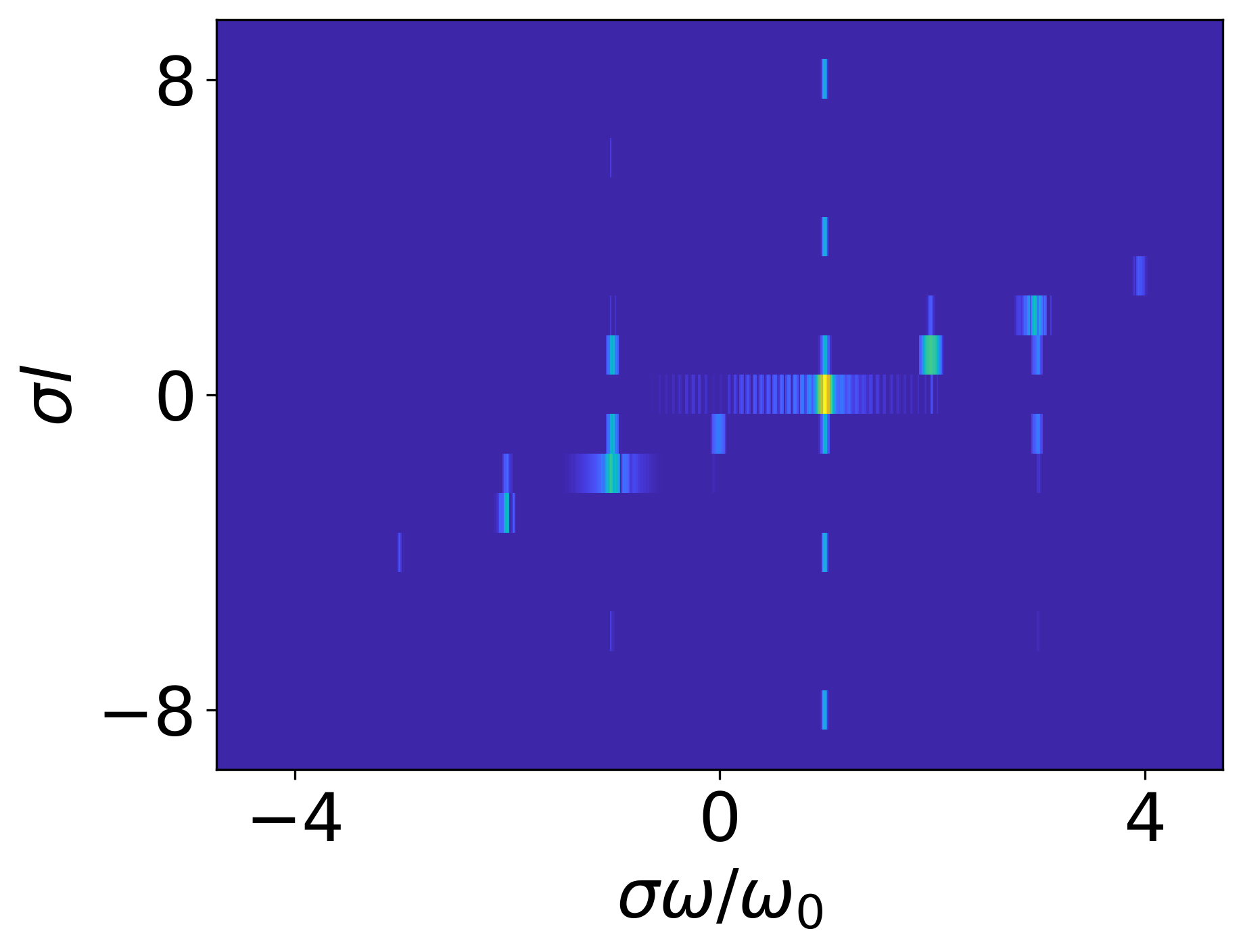}
        \includegraphics[width=0.3\columnwidth]{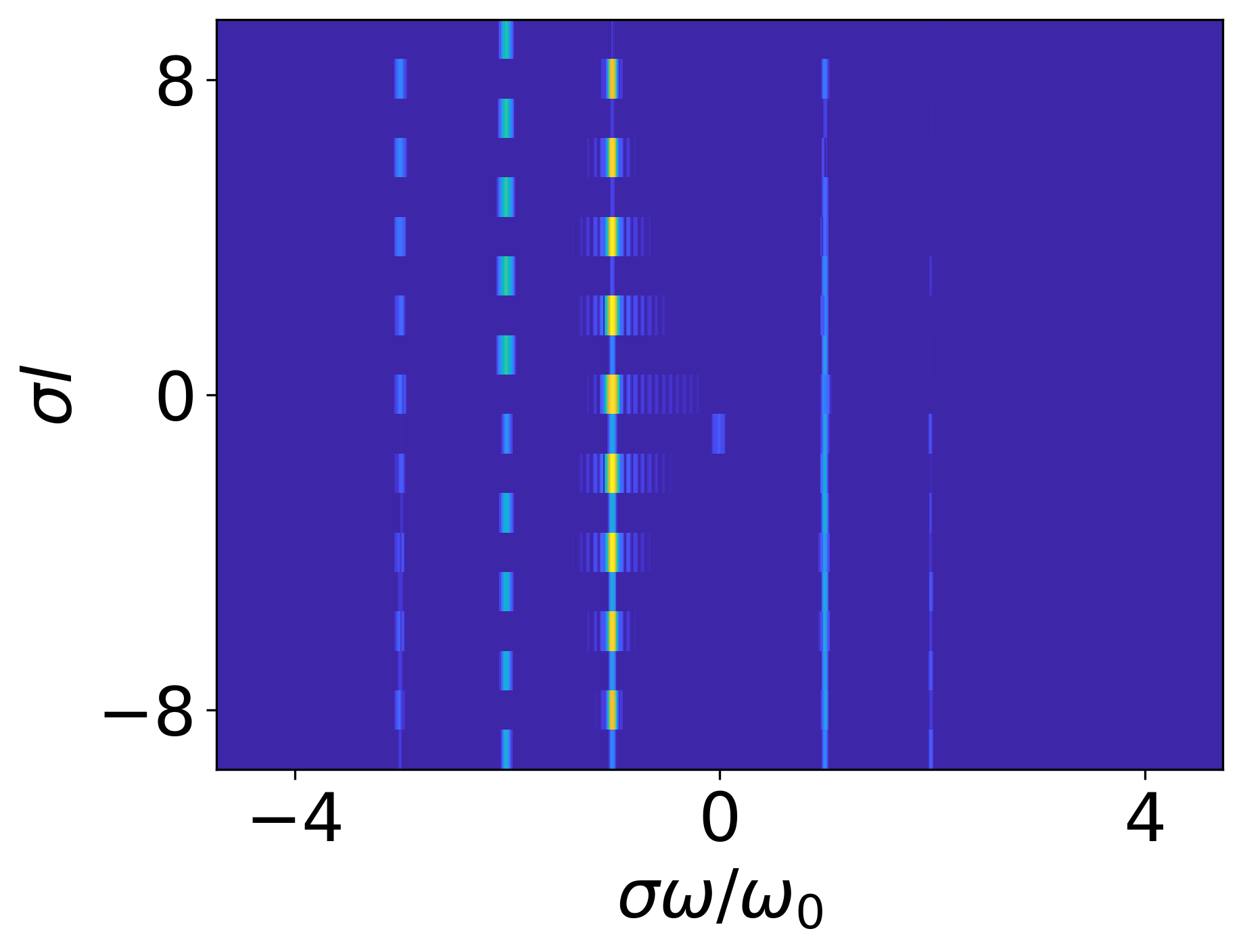}
        \includegraphics[width=0.3\columnwidth]{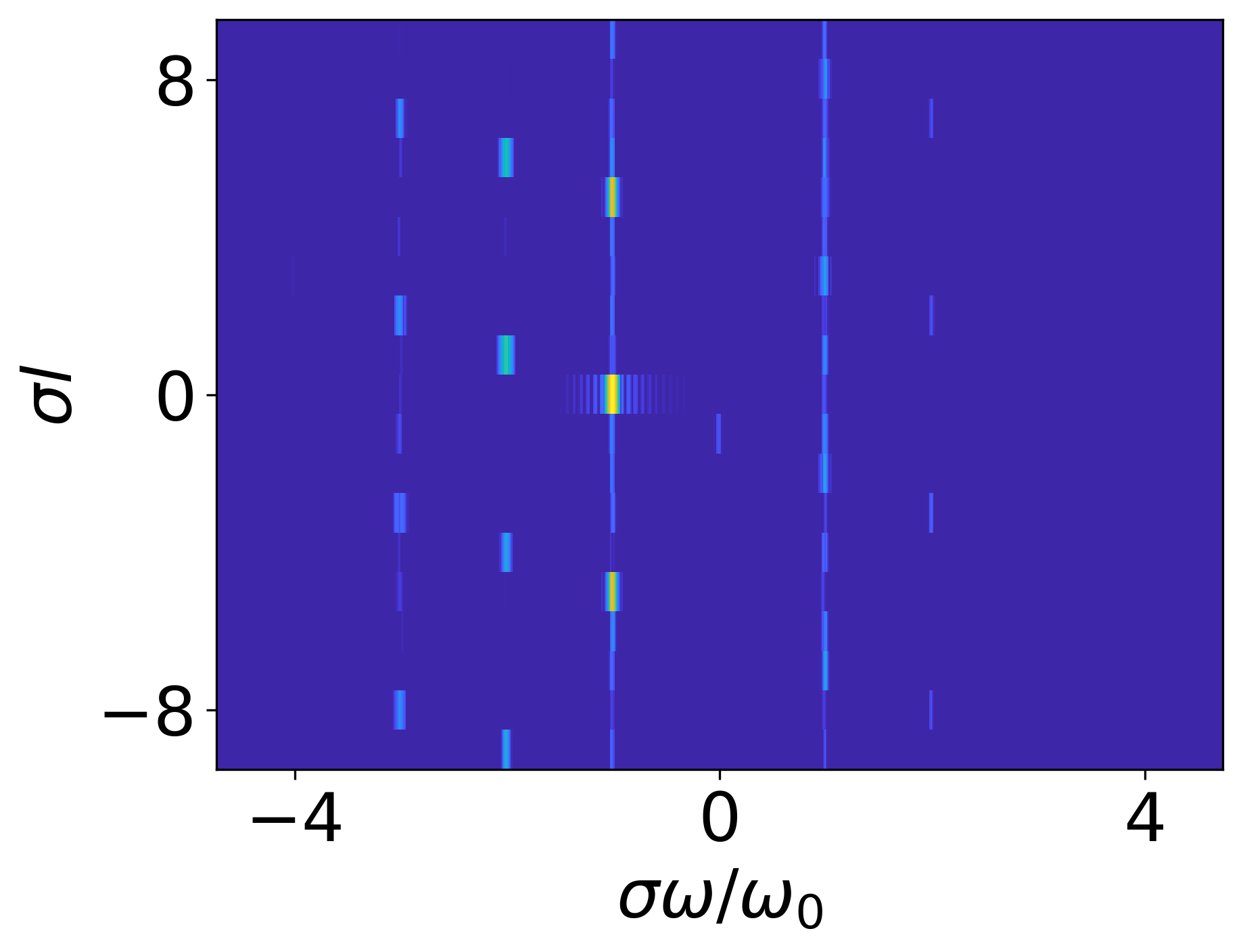}
    \end{center}
    
    \caption{Left: Spectrum $\sigma l$ vs $\sigma \omega$ resulting from a circularly polarised beam reflecting off a cone-shaped target. Middle, right: Spectrum $\sigma l$ vs $\sigma \omega$ resulting from a circularly polarised beam reflecting off a target with a structured surface with $q=-1$ (middle panel) and $q=-4$ (right panel).}
    \label{FigReflectionCircStructured}
\end{figure}

We start with the analysis for the simulations of ``reflective'' targets. The $(\sigma\omega,\sigma\ell)$ spectra of the reflected light are shown in Figure \ref{FigReflectionCircStructured}. In the left frame, we show the result for a CP pump laser beam with spin $\sigma_0=1$ and no OAM ($\ell_0=0$), corresponding to the point $(1,0)$. The cone-shaped dent leads to a DC mode with $\sigma\ell = -1$, as explained above, corresponding to the point $(0,-1)$. These two points are expected to define the harmonic progression $(\sigma\omega,\sigma\ell) = (n, n-1)$, which is precisely what was found in the 2-D Fourier spectrum from the simulation. We note in particular that we can also see ``negative'' harmonics with spin opposite to that of the pump beam; since their OAM level is clearly different from the corresponding ``positive'' harmonics, they are genuine harmonics and not numerical echoes. While these negative harmonics have been predicted by Li \emph{et al.} \cite{li2020}, they are actually demonstrated here for the first time. One can vary the values for $\sigma_0$ and $\ell_0$ so the pump laser corresponds to the point $(\sigma_0, \sigma_0 \ell_0)$, and this results in a harmonic progression $(n\sigma_0, n(1+\sigma_0 \ell_0)-1)$, defined by the points $(\sigma_0, \sigma_0 \ell_0)$ and $(0,-1)$, every time (not shown here).


Next, we replace the cone target with one with a structured surface corresponding to $q=-1$ and $q=-4$, corresponding to the points $(0,+1)$ and $(0,+4)$ respectively. The results are shown in the middle and right panels of Figure \ref{FigReflectionCircStructured}. We immediately observe that the point $(0,-1)$ is still present in both frames in addition to the new points  $(0,+1)$ or $(0,+4)$. Together with the point $(\sigma_0,\sigma_0\ell_0)$ from the pump laser, we now have three points not on a line in each frame, which define a 2-D grid of harmonic modes between them in both frames. We see that the vertical spacing of 2 and 5 respectively is controlled by the value of $q$ of the target. The ``slope'' of the rows of harmonic modes can be controlled by $\sigma_0$ and $\ell_0$ of the pump mode (not shown here). We note that modes with $\ell=0$ occur for only specific values of $\sigma\omega$; if one selects the on-axis harmonic light (where modes with $\ell=0$ dominate), one can obtain a controlled ``frequency comb'' \cite{alon98,bayku,saito17,rego22}: the vertical OAM spacing being converted into a horizontal frequency spacing (see also Section \ref{sec:design}). Note that for $\sigma_0\ell_0 = -1$, all harmonic modes in the spectrum will be in horizontal rows, so all frequencies exhibit the same OAM content and there will be no frequency comb.


\subsection{Why do we get a 2-D grid instead of a 1-D line?}
\label{sec:grid}

Let $h(x,y)$ be the ``height function'' of the structured target, and define $\Psi_h = \partial_x h + i\partial_y h$, while for the pump beam we write $\Psi_0 = E_x + iE_y$. Initially, the beating term appeared to be $\Psi_h^* \Psi_0$, but this ignores the phase shift induced by the variable target height. A better expression would be: $\exp[ 2i\sigma_0 k_0 h(x,y)] \Psi_h^* \Psi_0$, or similar. We introduce $h = h_0 (r/r0)^{1-q} \cos[(1-q)\varphi]$, and apply the Jacobi-Anger expansion to the complex exponential. Eventually, all this leads to a sequence of points $(0, -q + n(1-q))$, centred around $(0,-q)$ and always including the point $(0,-1)$. This explains why the harmonic progression for $q \not= 1$ is always 2-D, apparently generated by the points $(0,-1)$, $(0, -q)$ and $(\sigma,0)$.

For $q=-1$, $\nabla h$ only has a radial component, no azimuthal component. The resulting phase shift induced by the variable target height does not interfere with the OAM harmonic progression. This can be seen in Figure 2 (c) of Ref \cite{zepf}. For $q \not= -1$, $\nabla h$ does have an azimuthal component, and this does interfere with the OAM harmonic progression.

More generally, any target with $C_N$ symmetry (i.e. when the vector field $\nabla h$ or $\vec{n}$ is expressed in \emph{polar} coordinates, its $\vec{e}_r$ and $\vec{e}_\varphi$ coefficients will be periodic in $\varphi$ with period $2\pi/N$) will usually lead to points $(0,-1)$ and $(0,N-1)$. Such a target corresponds to a function $\Psi_{DC} \propto \exp(-i\sigma\ell\varphi)$ where $\Psi_{DC}\exp(-i\varphi)$ is $N$-periodic, so $\sigma\ell+1 = nN$, which includes the points $(0,-1)$ and $(0,N-1)$. We note that the structured target for a $q$-plate has $C_{|q|+1}$ symmetry. This will also come into play when studying the aperture targets below.

\subsection{Aperture simulations}

\begin{figure}[htb]
        \centering
        \includegraphics[width=1\columnwidth]{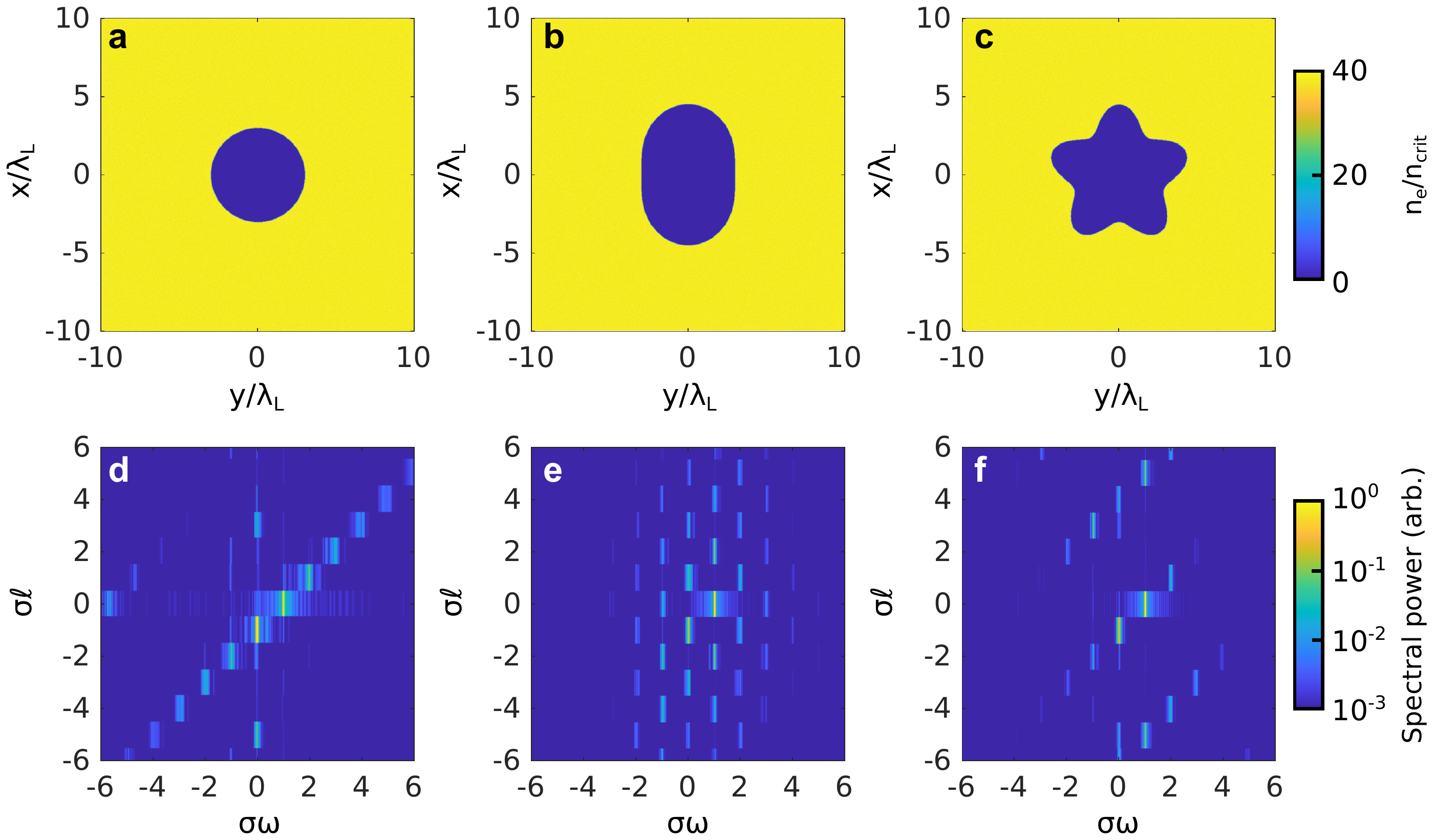}
        \caption{3D simulation results for the interaction of a circularly polarised pulse, $\sigma_0=1$, $\ell_0=0$, with an aperture target with a thickness $\lambda_{L}$ and three different periodic variations; (a-c) shows the initial electron densities for no periodic variation, $n=2$ and $n=5$ respectively at z=0.5$\lambda_L$, with the corresponding generated 2-D $\sigma\omega-\sigma \ell$ spectral map at $z=6\lambda_L$ (d-f).}
        \label{figureAp}
\end{figure}

Figure \ref{figureAp} (a-c) shows the initial electron densities in the $(x,y)$-plane for apertures with no periodic variation or periods of $n=2$ and $n=5$ respectively at $z=0.5$. The electric field components, $E_x$ and $E_y$ are sampled at the $(x,y)$-plane at $z=6\lambda_L$ every $0.06\tau_L$ and converted to cylindrical coordinates, such that $\Psi(r,\phi,t)=E_x+iE_y$. By integrating in $r$, and performing a 2-D Fourier transform, it is possible to obtain a 2D $\sigma\omega-\sigma l$ spectral map of the transmitted and generated light. This can be seen for the corresponding aperture periods of $n=0$, 2 and 5 in figure \ref{figureAp} (d-f), respectively.

For the aperture simulations, the normal vector of the inner target surface takes the role of the vector field $\nabla h$ in the simulations with reflective targets. For the purely circular target, the normal vector field is $-\vec{e}_r$, which has the same topology as $\nabla h$ for a reflective target with a cone-shaped dent. The targets with $n=2$ and $n=5$ have the same $C_N$ symmetry as the reflective targets for $q=-1$ and $q=-4$ respectively. We find that the 2-D $(\sigma\omega,\sigma\ell)$ spectra for the transmitted light of the aperture targets are identical to those of the reflected light of the reflective targets (apart from a change in $\sigma_0$). This makes good sense, since the topology of the respective cases was chosen to be the same in Figures \ref{FigReflectionCircStructured} and \ref{figureAp}.

Again, we note the presence of ``negative'' harmonics with spin opposite to that of the CP pump laser. These are simply ignored in Ref. \cite{yi21}, but are definitely present in the harmonic light and can be brought out via careful Fourier analysis of the function $\Psi = E_x + iE_y$. Especially the  negative harmonics at $\sigma\omega = -\omega_0$ would otherwise be drowned out by the transmitted light of the pump laser.


\section{Design alternatives}
\label{sec:design}

The design for the nonlinear q-plate can be extended in several directions.

(i) OAM may be added to the driving pump wave. For $q=1$, this can be done by using a target with a spiral surface, see e.g. Li \emph{et al.} \cite{li2020}. For $q \not= 1$, it may be difficult to add OAM via the structured target itself. A better option may be to add OAM to the pump beam via a separate optical element, see e.g. C. Brabetz \emph{et al.} \cite{brabetz}.

For $q \not= 1$, we find that the target is responsible for two points in $(\sigma\omega, \sigma\ell)$ space: $(0, -1)$ and $(0, -q)$, see Section \ref{sec:grid}. A CP laser beam then provides a third point: $(\sigma_0\omega_0,\sigma_0\ell_0)$. The harmonic progression driven by these three points is a 2-D grid with multiple OAM values per harmonic frequency, defined by the base vectors $(0,|1-q|)$ and $(\sigma_0\omega_0,1+\sigma_0\ell_0)$. For $\sigma_0\ell_0 = -1$, all harmonic frequencies will have the same OAM content, while for other values of $\sigma_0\ell_0$, the OAM content varies for adjacent harmonics, but is repeated with a period of up to $|1-q|$ harmonics.

For a 2-D grid where the OAM content varies between harmonic frequencies, selected harmonics will have an $\ell=0$ component with peak intensity on-axis, while the other harmonics without an $\ell=0$ component will have minimum intensity on-axis. When the on-axis harmonic light is then selected preferentially and analysed, a frequency comb is found \cite{rego22}, see also Refs. \cite{alon98,bayku,saito17}. The frequency comb is determined by finding the points $(\sigma\omega,0)$ in the 2-D harmonic progression (grid) in $(\sigma\omega, \sigma\ell)$ space. For laser beams without OAM interacting with structures with rotational symmetry, one finds $\sigma\omega = (nN+1)\sigma_0 \omega_0$ for the modes with $\sigma\ell = 0$ \cite{alon98,bayku,saito17}. As shown in Figures \ref{FigReflectionCircStructured} and \ref{figureAp}, we can generate a 2-D grid with the right properties for a laser hitting a structured target for judicious choices of $q$ and $\sigma_0 \ell_0$. The step size in the frequency comb is tuned via $N=|1-q|$, while the offset is tuned via $\sigma_0 \ell_0$.


(ii) The driving pump beam can have linear polarisation. When a non-linear q-plate is hit by a linearly polarised laser beam, the q-plate has one or two values for $\sigma\ell$, while the laser beam now has two values for $\sigma\ell$. This leads to a complex 2-D harmonic progression in $(\sigma\omega,\sigma\ell)$ space, with multiple OAM values for each harmonic frequency. See e.g. Bacon \textit{et al} \cite{Bacon2022}, where precisely this setup was used to study Hermite-Gaussian modes in the second harmonic light. One could compare this outcome to the 2-D $(\sigma\omega,\sigma\ell)$ spectrum found by Rego \emph{et al.} \cite{rego16,rego19}.

(iii) Zhang \emph{et al.} \cite{zhang21} use a tilted surface with a dent, effectively a superposition of $q=0$ and $q=1$ DC modes, with a circularly polarised laser beam, i.e. 3 modes in total. This again leads to a 2-D harmonic progression in $(\sigma\omega,\sigma\ell)$ space. One could extend this by using more complex superpositions of DC modes, to drive more complex harmonic progressions.

(iv) Qu, Jia and Fisch studied a plasma q-plate for $q=1$ using the magnetic field of two Helmholtz coils \cite{qplatequ}. We can now study plasma q-plates for $q \leq 0$ using the magnetic field of a magnetic $2(1-q)$ multipole (as used for particle beam optics), which already has the required shape. For $q<0$, it may be hard to generate a static magnetic field with the correct topology.

(v) The concept of the non-linear $q$-plate can be extended to that of
the non-linear J-plate \cite{jplate1,jplate2}. This works as follows. A basic
J-plate turns a circularly polarised beam with $\sigma=+1$ into one
with $\sigma=-1$, adding $r$ to the OAM level. Similarly, it turns a
circularly polarised beam with $\sigma=-1$ into $\sigma=+1$, adding
$s$ to the OAM level. To ensure that $r$ and $s$ are fully independent,
we need two degrees of freedom in the design of the non-linear
$q$-plate, which we choose to be $q$ and $\ell_s$. We revisit
Eq. (\ref{eq:qoam}), with nonzero $\ell_s$:
\[
\sigma_n\ell_n = n[\sigma(\ell+\ell_s) + q] -q.
\]
We set $n=-1$ and use $\sigma_{-1} = -\sigma$ to obtain:
\[
\ell_{-1} = \ell+\ell_s + 2\sigma q
\]
For $\sigma=+1$, we find $\ell_{-1} - \ell = r = \ell_s + 2q$, while
for $\sigma=-1$, we find $\ell_{-1} - \ell = s = \ell_s - 2q$. Thus,
a non-linear J-plate with parameters $r,s$ can be produced from
an enhanced non-linear $q$-plate by setting $q = (r-s)/4$ and
$\ell_s = (r+s)/2$.

\section{Conclusions and outlook}

In this paper, we have shown the following. We have developed a generic approach to high harmonic generation and harmonic progressions in $(\sigma\omega,\sigma\vec{k})$ space, in terms of the beating of ``fundamental modes'' with purely circular polarisation. We have demonstrated a novel way to analyse a harmonic $(\sigma\omega,\sigma\vec{k})$ spectrum via $\Psi = E_x + iE_y$ and the signed, multi-dimensional, complex Fourier transform of $\Psi$. We have reproduced earlier results on conical dent targets \cite{zepf,li2020} and aperture targets \cite{yi21}, and extended these results to show the presence of ``negative harmonics'' with polarisation opposite to that of the pump, which have not been seen before. We have demonstrated the generation of a rich two-dimensional harmonic spectrum from a single laser beam with circular polarisation, interacting with structured reflective and aperture targets. Finally, we have demonstrated that a two-dimensional harmonic progression in $(\sigma\omega,\sigma\ell)$ space can always be used to generate a tunable harmonic frequency comb, via selecting the harmonics $(\sigma\omega,0)$ from the 2-D spectrum. Now that we have revealed the underlying structure of many of these HHG processes, extensions in numerous directions are easily imagined. We have discussed several obvious examples, but envisage that there are many more.





\begin{thebibliography}{99}
\bibitem{zepf} J. W. Wang 1, M. Zepf and S.G. Rykovanov, Nature Commun. {\bf 10}, 5554 (2019).
\bibitem{li2020} S. Li \emph{et al.}, New J. Phys. {\bf 22}, 013054 (2020).
\bibitem{gariepy} G. Gariepy \emph{et al.}, Phys. Rev. Lett. {\bf 113}, 153901 (2014).
\bibitem{garcia13} C. Hern\'andez-Garc\'{\i}a \emph{et al.}, Phys. Rev. Lett. {\bf 111}, 083602 (2013).
\bibitem{rego16} L. Rego \emph{et al.}, Phys. Rev. Lett. {\bf 117}, 163202 (2016).
\bibitem{rego19} L. Rego \emph{et al.}, Science {\bf 364}, 1253 (2019).
\bibitem{torusknot} E. Pisanty \emph{et al.}, Phys. Rev. Lett. {\bf 122}, 203201 (2019).
\bibitem{rego22} L. Rego \emph{et al.}, Sci. Adv. {\bf 8}, eabj7380 (2022).
\bibitem{zhang15} X. Zhang \emph{et al.}, Phys. Rev. Lett. {\bf 114}, 173901 (2015).
\bibitem{zhang21} L. Zhang \emph{et al.}, Phys. Rev. Applied {\bf 16}, 014065 (2021).

\bibitem{GonzalezNatPhys2016} B. Gonzalez-Izquierdo, \emph{et al.}, Nat. Phys. {\bf 12}, 505 (2016).
\bibitem{GonzalezNatComm2016} B. Gonzalez-Izquierdo, \emph{et al.}, Nat. Comm. {\bf 7}, 12891 (2016).
\bibitem{DuffSciReports2020} M. J. Duff, \emph{et al.}, Sci. Rep. {\bf 10}, 105 (2020).
\bibitem{yi21} L. Yi, Phys. Rev. Lett. {\bf 126}, 134801 (2021).
\bibitem{Jirka2021} M. Jirka, O. Klimo, and M. Matys, Phys. Rev. Research {\bf 3}, 033175 (2021).
\bibitem{Bacon2022} E. F. J. Bacon, M. King, R. Wilson, T. P. Frazer, R. J. Gray, and P. McKenna, Matter and Radiation at Extremes {\bf 7}, 054401 (2022).



\bibitem{vieira2} J. Vieira \emph{et al.}, Phys. Rev. Lett. {\bf 117}, 265001 (2016).
\bibitem{beat1} C. Joshi, T. Tajima, J. M. Dawson, H. A. Baldis, and N. A. Ebrahim, Phys. Rev. Lett. {\bf 47}, 1285 (1981).
\bibitem{beat2} R. Bingham, Physica Scripta {\bf T30}, 24, (1990).
\bibitem{bloembergen80} N. Bloembergen, J. Opt. Soc. Am. {\bf 70}, 1429 (1980).
\bibitem{schafer93} K. J. Schafer \emph{et al.}, Phys. Rev. Lett. {\bf 70}, 1599 (1993). 
\bibitem{corkum93} P. B. Corkum, Phys. Rev. Lett. {\bf 71}, 1994 (1993).
\bibitem{nonper2} M. Lewenstein, \emph{et al.}, Phys. Rev. A {\bf 49}, 2117 (1994).
\bibitem{nonper1} Anne L'Huillier and Philippe Balcou, Phys. Rev. A {\bf 46}, 2778 (1992).
\bibitem{long95} S. Long, W. Becker, and J. K. McIver, Phys. Rev. A {\bf 52}, 2262 (1995).
\bibitem{nonper3} D. B. Milo\v{s}evi\'c, W. Becker and R. Kopold, Phys. Rev. A {\bf 61}, 063403 (2000).

\bibitem{fleischer} A. Fleischer \emph{et al.}, Nat. Photonics {\bf 8}, 543 (2014).
\bibitem{kfir2} O. Kfir \emph{et al.}, Nature Photonics {\bf 9}, 99 (2015).
\bibitem{lichters} R. Lichters, J. Meyer-ter-Vehn, and A. Pukhov, Phys. Plasmas {\bf 3}, 3425 (1996).
\bibitem{koba12} H. Kobayashi, K. Nonaka, and M. Kitano, Opt. Expr. {\bf 20}, 
14064 (2012).
\bibitem{huang20} Chen-Kang Huang \emph{et al.}, Comm. Phys. {\bf 3}, 213 (2020).
\bibitem{qplate1} L. Marrucci,  C. Manzo and D. Paparo, Phys. Rev. Lett. {\bf 96}, 163905 (2006).
\bibitem{qplate2} L. Marucci \emph{et al.}, J. Opt. {\bf 13} 064001 (2011).
\bibitem{maki95} J. J. Maki, M. Kauranen and A. Persoons. Phys. Rev. B {\bf 51}, 1425 (1995).
\bibitem{courtial98} J. Courtial \emph{et al.}, Phys. Rev. Lett. {\bf 81}, 4828 (1998).
\bibitem{Arber:2015} T. D. Arber \emph{et al.}, Plasma Phys. Control. Fusion {\bf 57}, 113001 (2015).
\bibitem{brabetz} C. Brabetz \emph{et al.}, Phys. Plasmas {\bf 22}, 013105 (2015).
\bibitem{alon98} O. E. Alon, V. Averbukh and N. Moiseyev, Phys. Rev. Lett. {\bf 80}, 3743 (1998).
\bibitem{bayku} D. Baykusheva \emph{et al.}, Phys. Rev. A {\bf 103}, 023101 (2021).
\bibitem{saito17} N. Saito \emph{et al.}, Optica 4, {\bf 1333}, (2017).
\bibitem{qplatequ} K. Qu, Q. Jia and N. J. Fisch, Phys. Rev. E {\bf 96}, 053207 (2017).
\bibitem{jplate1} R. C. Devlin \emph{et al.}, Science {\bf 358}, 896 (2017).
\bibitem{jplate2} H. Sroor \emph{et al.}, Nature Photonics {\bf 14}, 498 (2020). 


\end{thebibliography}
\end{document}